\documentclass[12pt,preprint]{aastex}

\begin{document}
\title{Damped Ly$\alpha$ Absorbing Galaxies At Low Redshifts $\small{z} \le
1$ From Hierarchical Galaxy Formation Models}

\author{Katsuya Okoshi}
\affil{National Astronomical Observatory,
Mitaka,  Tokyo 181-8588,  Japan;
\email{okoshi.katsuya@nao.ac.jp}}
\and
\author{Masahiro Nagashima}
\affil{Department of Physics, University of
Durham, South Road, Durham DH1 3LE, England\\
Department of Physics, Graduate School of Science, Kyoto University,
Sakyo-ku, Kyoto 606-8502, Japan
}

\begin{abstract}

We investigate Damped Ly$\alpha$ absorbing galaxies (DLA galaxies) at
low redshifts $z \le 1$ in the hierarchical structure formation
scenario.  In our previous paper, we showed that our model of galaxy
formation can explain basic properties of Damped Ly$\alpha$ (DLA)
systems such as the metallicity evolution and \ion{H}{1} column density
distribution.  As a subsequent study, we focus on low-redshift DLA
systems in detail to be compared with recent data of many
characteristics and their relationships such as luminosities, \ion{H}{1}
column densities and sizes of galaxies with large \ion{H}{1} column
densities enough to produce DLA absorptions obtained by optical/radio
observations.  While it has been debated about what types of galaxies
correspond to DLA systems, by using a theoretical model simultaneously
treating with both DLA systems and galaxies, we clarify the nature of
low-redshift galaxies producing DLA absorptions because observational
data of such galaxies mainly at low redshifts are currently available.
We find that our model well reproduces distributions of fundamental
properties of DLA galaxies such as luminosities, column densities,
impact parameters obtained by optical and near-infrared imagings.  Our
results suggest that DLA systems primarily consist of low luminosity
galaxies with small impact parameters (typical radius $\sim 3$ kpc,
surface brightness from $22$ to $27$ mag arcsec$^{-2}$ ) similar to low
surface brightness (LSB) galaxies.  In addition, we investigate
selection biases arising from the faintness and from {\it the masking
effect} which prevents us from identifying a DLA galaxy hidden or
contaminated by a point spread function of a background quasar.  We find
that the latter affects the distributions of DLA properties more
seriously rather than the former, and that the observational data are
well reproduced only when taking into account the masking effect.  The
missing rate of DLA galaxies by the masking effect attains $60-90 \%$ in
the sample at redshift $0 \le z \le 1$ when an angular size limit is as
small as 1 arcsec.  Furthermore we find a tight correlation between
\ion{H}{1} mass and cross section of DLA galaxies, and also find that
\ion{H}{1}-rich galaxies with M$_{\rm HI} \sim 10^{9}$ M$_{\odot}$
dominate DLA systems at $z \sim 0$. These features are entirely
consistent with those from the Arecibo Dual-Beam Survey which is a blind
$21$ cm survey. Finally we discuss star formation rates, and find that
they are typically about $10^{-2}$ M$_{\odot}$ yr$^{-1}$ as low as those
in LSB galaxies.

\end{abstract}



\section{Introduction} 

Numerous absorption lines found in the quasar spectra 
are one of few observational opportunities that provide us fruitful
information on the physical state of the evolving universe. 
These absorption lines offer several advantages over emission lines. 
For example, the line features reflect the physical state of the
astronomical objects and the intergalactic medium.  Another advantage
comes from the fact that absorption lines are free from observational 
limitations of the detection of absorbers caused by their faintness 
in photometric surveys.

Among those absorption line systems, damped Ly-alpha (DLA) systems would
provide exceptional insights into exploring galaxy formation.  DLA
systems have been interpreted to arise from cold gas in galactic disks
along lines of sight to quasars \citep{WTSC}.  So far some observational
facts on DLA systems are obtained from detailed studies of quasar
absorption spectra, for example, (1) the \ion{H}{1} column densities are
similar to those in our Galaxy, (2) most of them have low metallicities,
$\sim 1/10 Z_{\odot}$ \citep[e.g.][]{P03} and (3) the \ion{H}{1} column
density distributions are fitted by a single power law as well as those
of the Ly$\alpha$ forest \citep[e.g.][]{SW00, Peroux03}.  These facts
suggest that DLA systems are galaxies in an early evolutionary
stage. Therefore, it could be an interesting issue whether theoretical
models of galaxy formation can consistently account for fundamental
properties of DLA systems.

Furthermore, some low-redshift DLA systems can be also directly observed
in recent photometric surveys. These samples provide a clue to revealing
what types of intervening galaxies arise DLA lines in quasar spectra,
which are referred as `DLA galaxies' hereafter. Such low-redshift DLA
galaxies have been studied from both direct photometric images and
spectroscopic follow-ups (e.g. Steidel et al. 1994, 1997; Lanzetta et
al. 1997; Le Brun et al. 1997; Fynbo et al. 1999; Rao \& Turnshek 1999;
Turnshek et al. 2001; Bouch$\rm \acute{e}$ et al. 2001; Bowen, Tripp \&
Jenkins 2001; Warren et al. 2001; M{\o}ller et al. 2002; Rao et
al. 2003; Chen \& Lanzetta 2003; Schulte-Ladbeck et al. 2004; M{\o}ller,
Fynbo \& Fall 2004).  The results of the searches suggest that DLA
systems have a wide range of
the morphology from dwarf galaxies to spirals and do not comprise a
single population such as normal spiral galaxies.  This picture is
rather consistent with what is expected from the hierarchical structure
formation scenario based on a cold dark matter (CDM) models because
those predict that galaxies can span a mixture of their morphological
types from dwarf to massive spiral galaxies. So it is clearly valuable
to compare the observed properties of DLA galaxies with those predicted
by theoretical models as a useful test of theories of galaxy formation.

Semi-analytic modeling has been applied with a view to deciphering the
clues to the formation process of galaxies in the hierarchical
clustering scenario.  This approach takes into account merging histories
of dark halos based on the power spectrum of the initial density
fluctuation, and has successfully provided galaxy formation models for
explaining observational properties of galaxies such as luminosity
functions, the relation between \ion{H}{1} gas mass fraction and
luminosities, and so forth.  It has some advantages over numerical
hydrodynamical simulations. For example, it can clarify the effect of
star formation or supernovae feedback on galaxy evolution even under
simple recipes. Moreover, it does not also suffer from resolution
limitations in numerical hydrodynamic simulations. This is important to
study the formation process of small objects that are hardly resolved by
numerical simulations.  Therefore, it is valuable to apply this model
for studying the evolution of DLA systems which tightly correlate with
galaxies.  So far, several semi-analytic models have been developed and
provided interesting results for physical relations between DLA systems
and galaxies \citep[][hereafter Paper I]{K96, spf01, MPSP01,MPSP03,
ONGY}.  In Paper I, we focused on the metallicity evolution and the
\ion{H}{1} column density distribution, and concluded that DLA systems
primarily consist of dwarf and/or sub-$L^{*}$ galaxies. As a subsequent
study, this paper expands on these previous results to reveal the nature
of low-redshift DLA systems by exploring typical properties of DLA
galaxies obtained by recent observations. Here, the following advantages
of this study should be addressed. (1) Our model can reproduce main
properties of DLA systems, that is, the metallicity evolution and the
\ion{H}{1} column density distribution, (2) Our model incorporates with
various effects in detecting DLA galaxies: cosmological dimming of the
surface brightness, internal dust absorption and the observational bias
caused by glare of quasars behind DLA galaxies along lines of sight. (3)
Our model can also reproduce many observational results of galaxy
population such as luminosity functions and number counts
\citep[see][]{NTGY}. This is the first theoretical study using a
hierarchical galaxy formation model to explore photometric and radio
properties of low-redshift DLA galaxies comprehensively enough to
compare with the currently available observations.

In \S 2, we briefly describe our model. In \S 3, we show the results for
various properties of DLA galaxies.  In \S 4, we discuss selection
biases for detection of DLA galaxies. In \S 5, we explore some
possibilities to study the nature of DLA galaxies. We focus on the radio
properties of DLA systems.  We also discuss the star formation rates in
DLA galaxies, which can be a good tracer for discerning what types of
galaxies comprise the population of DLA systems. Finally we summarize
our conclusions and discuss our results in comparison with other
observations in \S 6.


\section{Model}

The semi-analytic model of galaxy formation used here is based on cold
dark matter models in which, assuming a power spectrum of initial
density fluctuations, dark halos emerge from the density fluctuations,
cluster gravitationally, and merge together. Gas in dark halos cools and
then forms stars. Such processes lead to the formation of galaxies which
comprise gas and stars embedded in dark halos. After galaxies form, some
galaxies grow up to massive ones like our Galaxy via merging
processes. We apply this scenario to the study of the nature and
formation of DLA systems in a framework of semi-analytic modeling. The
semi-analytic model of galaxy formation used here is the same as the LC
model described by Paper I, which well reproduces the observed number
distributions and metallicity evolution of DLA systems.  This model also
well reproduces many aspects of observed properties of galaxies such as
luminosity functions, cold gas mass fractions, disk sizes, and faint
galaxy number counts in a Lambda-CDM model (Nagashima et al. 2001).  We
provide an outline of this model below.
 
The cosmological parameters we adopted are $\Omega_{0}=0.3$,
$\Omega_{\Lambda}=0.7$, $\Omega_{\rm b}=0.015h^{-2}$, $h=0.7$ (where $h$
is the Hubble parameter, $h=H_{0}/100$km~s$^{-1}$ Mpc$^{-1}$), and
$\sigma_{8}=1$ that is the normalization of the power spectrum of
density fluctuations given by \citet{BBKS}.  The number density of dark
halos at present is given by the Press-Schechter mass function
\citep{PS74}.  The past merging history of each dark halo is realized by
a Monte Carlo method proposed by \citet{SK99}, which is based on an
extended Press-Schechter formalism \citep{BCEK, B91, LC93}.  Only halos
with circular velocities $V_{\rm circ}\geq 40$ km s$^{-1}$ are
identified as isolated halos, and others are regarded as the diffuse
accretion mass.

We assume that baryonic gas consists of two phases: cold and hot.  The
gas in a halo should be shock-heated to the the virial temperature of
the halo after the halo collapses. The heated gas is defined as hot gas
in our model. A part of hot gas cools quickly by radiative cooling and
falls to gaseous disks, which is defined as cold gas. The cold gas then
becomes available for star formation.  The star formation rate (SFR) is
assumed as
\begin{eqnarray}
\dot{M}_{*}=\frac{\displaystyle M_{\rm cold}}{\displaystyle 
\tau_{*}}, 
\end{eqnarray}
where $M_{*}$ and $M_{\rm cold}$ are the masses in stars and cold gas,
respectively, and $\tau_{*}$ is the timescale of star formation.  We
assume a star formation timescale independent of redshift as follows,
\begin{eqnarray}
\tau_{*}=
\displaystyle{\tau_{*}^{0}\left(\frac{V_{\rm circ}}{300~{\rm km/s}}\right)
^{\alpha_{*}}} .
\end{eqnarray}
The free parameters $\tau_{*}^{0}$ and $\alpha_{*}$ are chosen by
matching the model prediction of cold gas mass fractions of spiral
galaxies to observed one because those directly determine the gas
consumption rate.  Therefore, they should play an important role in
determining the observable characteristics of DLA systems.  As a result
in Paper I, we have found that the above prescription of star formation
with $(\tau_{*}^{0}, \alpha_{*})$ $=$ ($1.5\mbox{Gyr}$, $-2$)
successfully reproduces both metallicity evolution and \ion{H}{1} column
density distributions of observed DLA systems (see Figures 1 and 3 in
Paper I). Thus, we adopt these values in this study. From the estimated
SFR, luminosities of galaxies are computed by using simple stellar
populations given by \citet{KA97}. We also include a supernova feedback
process and merging process of galaxies. The details are described in
Paper I.

Finally, we address DLA systems in our model. We simply assume that all
DLA systems have gaseous disks which are face-on to an observer because
the inclination effect hardly affects the column density distributions
(see Figure 4 in Paper I). Here, the radial distribution of the
\ion{H}{1} column density follows an exponential profile with an
effective radius of a gaseous disk, $r_{\rm e}$. It is assumed to be
$r_{\rm e}=r_{0} (1+z)$ where $r_{0}$ is a radius provided by the
specific angular momentum conservation of cooling hot gas. We also
assume that the dimensionless spin parameter has a log-normal
distribution with the average $0.06$ and the logarithmic variance $0.6$.
The central column density $N_{\rm 0}$ is given by $N_{\rm 0}=M_{\rm
cold}/(2 \pi \mu m_{\rm H} r_{\rm e}^{2})$, where $m_{\rm H}$ is the
mass of a hydrogen atom and $\mu(=1.3)$ is the mean molecular
weight. The size of a DLA system is defined by the radius $R$ at which
$N_{\rm HI} = 10^{20}$ cm$^{-2}$. For each system, we take the column
density averaged over radius within $R$. The above definition of DLA
systems is the same as in Paper I.

We assume that cold gas in DLA systems is neutral.  This would be
justified as follows.  The \ion{H}{1} column density of DLA systems
exceeds about $10^{20}$ cm$^{-2}$, which means that the cold gas is
optically thick.  So the ionization fraction averaged over the whole
disk should be very small even if the UV background radiation exists
around DLA systems.  \citet{PW96} calculated the ionization fraction in
DLA systems taking into account radiation transfer, assuming a UV
background intensity corresponding to that at $z \sim 2-3$, which
probably be higher than at $z \la 1$.  When a DLA system is a uniform
gas layer with a number density $n=0.1$ cm$^{-3}$, which is similar to
that of typical DLA systems in our model, they found the ionization
fraction $x(=n_{\rm e}/n) < 0.1$.  It could be also possible that far UV
radiation emitted from internal OB stars ionizes the hydrogen atoms in
disks.  Radio observations of ionized gas in our Galactic disk, however,
have revealed that the mass ratio of \ion{H}{2} to \ion{H}{1} gases is
about $0.01$, and that the filling factor of \ion{H}{1} gas is less than
$0.1$ \citep{O89}.  We consider that the ionization fraction in DLA
systems should be smaller than that in our Galaxy because ionizing
photons are expected to be less than in our Galaxy, inferred from
smaller SFRs in DLA systems.  This feature has been also confirmed by
recent measurements of the \ion{Al}{3} abundance in DLA systems
\citep{V01}. The \ion{Al}{3} abundance is a good tracer for estimating
the intensity of ionizing UV radiation because the production of
\ion{Al}{3} requires UV photons. They concluded that cold gas in DLA
systems is almost neutral from the observed small abundance of
\ion{Al}{3}. Therefore, it is reasonable to assume that cold gas in DLA
systems is neutral.

\section{DLA Galaxy Properties at low redshift}

As shown in Paper I, our model has a good ability to reproduce various
properties not only of galaxies both at high and low redshifts but of
DLA systems, particularly the observed distributions in the \ion{H}{1}
column density and the metallicity evolution.  Here we present various
properties of DLA galaxies at redshift $0 \le z \le 1$.  Recently,
\citet{R03} compiled observational data of 14 DLA systems, including new
identified ones, at $0 \le z \le 1$ and showed distributions of their
properties (luminosities, neutral hydrogen column densities, impact
parameters and number distributions).  Their conclusion is that
low-luminosity dwarf galaxies with small impact parameters dominate
their compiled sample.  Below we compare statistical properties of DLA
systems in our model with the observations.  Note that, in our
calculation, the total number of model DLA galaxies is very large, for
example, $\sim 8 \times 10^{3}$ and the covering comoving volume is
$\sim 2 \times 10^{9}$ Mpc$^{3}$ at redshift $z=0$, which are
statistically large enough to investigate the DLA properties.  Note also
that when presenting averaged values over redshifts, the comoving volume
element $dV/dz$ is taken into account.

Following the presentation of the observational data by \citet{R03}, 
Figure 1 shows distributions of properties of DLA galaxies about
luminosities in $B$-band relative to $L^{*}$, which corresponds to
$M_{\rm B}=-20.9$ mag \citep{R03}, neutral hydrogen column densities
$N_{\rm HI}$ and radii $b$.  Figure 1(a) shows the luminosity evolution
from $z=1$ to present.  We find that average luminosities predicted by
our model are broadly consistent with those of some observed DLA
galaxies at low redshifts, while some are brighter than our results.  As
shown in Paper I, our model predicts that the average circular
velocities of dark halos hosting DLA systems increase toward low
redshift as merging proceeds, and that the average attains $V_{\rm circ}
\sim 90$ km s$^{-1}$ at redshift $z=0$ (see Figure 8 in Paper I). The
luminosities also increase gradually as the star formation proceeds, and
the average luminosity is $L \sim 2 \times 10^{9} L_{\odot} \sim
0.05L^{*}$ at present (see Figure 10 in Paper I), which are fainter than
those of some normal spirals, $ \sim L^{*}$.  Thus our model apparently
predict the average luminosities of DLA galaxies lower than the
observations, which include some $L_{*}$-galaxies as shown in Figure
1(a), unless any relevant selection biases as discussed in next section
are taken into account.

In Figures 1(b) and (c), our results show that DLA galaxies typically
have neutral hydrogen column densities $N_{\rm HI} \sim 10^{20.6} $
cm$^{-2}$ and radii $b \sim 3$ kpc (see also panels i and j), and their
evolution is very moderate at $z\la 1$.  In Figure 1(c), it appears that
the mean sizes are generally smaller than the observations. Because the
radial sizes in our calculation provide upper limits of impact
parameters, our model seems to underpredict the sizes of DLA
systems. Figure 1(d) depicts the number fraction of DLA galaxies as a
function of redshift.  Our result shows that the number increases toward
higher redshift. This matches the observational trend of the redshift
distribution of {\it absorption line} systems, $dN/dz$, that their
number increases up to $z \sim 5$ \citep[e.g.][]{SW00, PH04}. Note that
the number of DLA galaxies in panel(d) is not identical to $dN/dz$.  The
former is weighted by the comoving volume element $dV/dz$ and the latter
by the cross section of DLA systems.

Figures 1(e) and (f) present the mean column density and impact
parameter as a function of luminosity.  DLA galaxies with lower
luminosities tend to have smaller impact parameters, while {\it mean}
column densities depend very weakly on the luminosities for $L/L^{*} \ga
0.1$.  Note that Figure 1(f) indicates that some bright galaxies have
large sizes with radii $ b \ga 10$ kpc while the mean is $\sim 3$ kpc.
Our result also suggests that DLA galaxies with large $b$ ($\ga 20$ kpc)
should be identified as $L^{*}$ spirals. We thus emphasize that luminous
galaxies ($\ga L^{*}$) arise damped Ly$\alpha$ absorptions although the
number fraction is much lower than that of the dominant population ($L
\la 0.1 L^{*}$).  This is in agreement with the observational fact that
DLA systems have a wide range of morphology from dwarf galaxies to
bright spirals.

We also show how the extent of neutral gas around DLA galaxies scales
with the column densities in Figure 1(h).  The relation between $b$ and
$\log N_{\rm HI}$ suggests that most of DLA systems have impact
parameters smaller than $10$ kpc.  This figure also shows a trend that
DLA galaxies with large $N_{\rm HI}$ have small sizes, similar to the
observation.  Our model, however, seems to underpredict the size as
shown in Figure 1(c). This reason is discussed in next section.

From the above results, it is suggested that DLA systems mainly comprise
dwarf galaxies with small sizes.  This is consistent with some trends
emerging in the observation \citep{R03}. Nevertheless, some
observational data show that DLA galaxies are likely to be brighter and
observed at impact parameters larger than our results although the
available sample is still small in the observations.  These differences
might be also alleviated if some selection biases exist such that bright
galaxies should likely be observed with large impact parameters, in
other words, most of optical counterparts of DLA systems are hardly
identified because of the faintness.  In general, the identification of
DLA galaxies among a large number of the candidates requires both
photometric images and spectroscopic follow-ups.  It is often extremely
difficult to pick out their candidates and/or to identify them
accurately. This is partly because they have low surface brightness and
partly because the candidates exist in close proximity to a quasar line
of sight.  Below we investigate how such selection biases affect the
distributions of DLA-galaxy properties discussed here.

\section{Selection bias}

\subsection{Surface Brightness Limit}

Firstly, we focus on surface brightness distributions of DLA galaxies at
redshift $0 \le z \le 1$ because detection limits are usually determined
by a surface brightness.  Figure 2 shows the $B$-band central surface
brightness $\mu$ of DLA galaxies.  Figure 2(a) depicts the evolution of
their mean surface brightness.  We find that the surface brightness
becomes fainter toward higher redshift and that the mean surface
brightnesses are $\sim 22$ mag arcsec$^{-2}$ at $z \sim 0$ and $\sim 27$
mag arcsec$^{-2}$ at $z \sim 1$.  The evolution of surface brightnesses
arises from both the luminosity evolution and cosmological dimming.

We also show distributions of central surface brightness averaged over
$0 \le z \le 1$ weighted by the comoving volume element against
luminosities, column densities and sizes, in Figures 2 (b)-(d),
respectively.  Our results suggest that the surface brightness does not
apparently depend on both the luminosity for $L/L^{*} \ga 0.1$ and the
radius but the column density.

Figure 3 shows the number fractions as a function of $\mu$ at redshifts
$z=0,0.5$ and $1$.  These results clearly represent that DLA galaxies
become brighter toward low redshift and the mean central surface
brightness does by about 5 mag from $z=1$ to $0$.  We also find the
average surface brightness $\mu \sim 25$ mag arcsec$^{-2}$ at $0 \le z
\le 1$.

Very low surface brightness DLA galaxies computed here are not expected 
to be detected because of the faintness below the detection limit.
Although the limit depends on the quasar fields, for example,
\citet{R03} set their $3 \sigma$ limiting surface brightnesses of
around $25$ mag arcsec$^{-2}$ (Table 2 in their paper).  According to
the surface brightness limit, we pick out DLA systems with $\mu \le 25$
mag arcsec$^{-2}$ and show distributions of their properties in Figures
4, in which the number fractions shown in bottom panels are defined as
the ratios the number of DLA galaxies per bin to that of DLA galaxies
which fulfill the selection criteria.  Our results show that most
distributions are very similar to those without the limit in Figure 1,
while the mean $N_{\rm HI}$ becomes larger particularly at high redshift
$z \sim 1$ and the number fraction in $N_{\rm HI}$ (panel i) shows that
DLA galaxies apparently decrease in low $N_{\rm HI} \la 10^{21}$
cm$^{-2}$.  These tendencies stem from the following reasons.  In
Figures 2, we found that high surface brightness galaxies have gaseous
disks with high $N_{\rm HI}$. In other words, the surface brightness is
strongly correlated with the column density.  At high redshifts, the low
surface brightness galaxies dominate the population of DLA systems.
Therefore, picking out only the high surface brightness systems arises
the increase of the mean column densities at high redshifts.  In
contrast, the surface brightness is not significantly correlated with
the luminosity and the size.  This causes the results that the
distributions of $L$ and $b$ are still similar to those without the
surface brightness limit.  Therefore, we find that this selection effect
does not account for the observed distributions of DLA galaxy properties
when the limit is as low as $\mu \sim 25$ mag arcsec$^{-2}$.

\subsection{Angular Size Limit}

Second, we focus on a selection effect caused by angular sizes of DLA
galaxies.  Identification of a DLA galaxy requires accurate
determination of its redshift which must be identical to that of a DLA
system as a possible counterpart.  If a DLA galaxy is compact and/or
very close to a quasar line of sight, it often happens that the image is
blended or hidden in the point spread function (PSF) of the background
quasar.  Even after subtracting the PSF from the blended image, noisy
residuals often remain so that any information cannot be extracted
within the radius of the circle enclosed by the PSF.  {\it If a galaxy
that gives arise to damped Ly$\alpha$ absorption is small and comparable
to the size of the PSF}, it is more likely to arise this difficulty
because the DLA galaxy is contaminated or hidden by the PSF.  Even when
the size is $\sim 1$ arcsec, it corresponds to the scale at most $\sim
6$ kpc at $z=0.5$ and $\sim 8$ kpc at $z=1$ in the cosmological model
adopted here.  This selection effect, which is called the `masking
effect' hereafter, can be serious because DLA galaxies are expected to
be very compact, about 3kpc on average in our model.

Figure 5 presents the distributions of DLA galaxies as a function of the
angular radius $\theta$ at $z=0.1$ ({\it dotted line}), 0.5 ({\it dashed
line}), and 1 ({\it solid line}), respectively.  The number fraction in
large $\theta$ decreases toward high redshift while the physical size of
DLA galaxies apparently shows no evolution [see Figure 1(c)].  Thus, the
evolution of the number distribution with redshift should be caused just
by the distance varying which directly affects the angular size.

In our model, $60-90 \% $ of DLA systems have small angular sizes less
than 1 arcsec.  The limits of angular sizes caused by the PSF are
usually different in observed images by images.  In most observations,
however, the limiting angular sizes are likely to be larger than 1
arcsec.  Here, we set the angular size limit $\theta_{\rm th} = 1$
arcsec and assume that DLA systems with angular sizes smaller than 1 
arcsec are not precisely resolved or identified as DLA galaxies by the 
masking effect.

Taking into account this selection bias in addition to the surface
brightness limit $\mu_{\rm th}=25$ mag arcsec$^{-2}$, we calculate
distributions of DLA galaxies with angular sizes larger than 1
arcsec. The results are presented in Figure 6.  We find that, at high
redshifts, DLA systems with large sizes mainly contribute to optically
observable DLA galaxies because the physical size corresponding to
$\theta = 1$ arcsec increases toward high redshifts.  This is confirmed
by the result for the evolution of their sizes in Figure 6(c) in which
the averages are larger than those in Figure 1(c) at $z \ga 0.5$.
Figure 6(c) suggests that DLA galaxies with large radii, $\sim 10$ kpc,
obviously increase at redshift $z \ga 0.5$ under the masking effect
(cf. Figure 1(c)).  It also appears that the number fraction of large
$b$ systems averaged over $0 \leq z \leq 1$ increases as shown in Figure
6(j) in comparison with Figure 1(j).  The masking effect evidently causes
better agreement with the observational data of the impact parameters
particularly for large $b$ [Figures 6(c) and (h)].  Figure 6(a) presents
that this bias also affects the apparent evolution of the luminosity of
observed DLA galaxies.  The luminosities agree better with the
observations in Figure 6(a) by taking into account the masking effect.
This is because most of large DLA systems, which are dominant at high
redshift, are bright as shown in Figure 6(f).  Figure 6(d) shows that
the redshift distribution of DLA galaxies exhibits a trend similar to 
that in Figure 1(d) for a model without the selection effects, while 
the number evolution becomes milder.

Figure 6 is one of the main results of this study, that is, it shows
that the selection biases, particularly the masking effect, are very
important when we interpret the observed results of DLA galaxies, and
that the biases lead to much better agreement with the observations.
This suggests that bright DLA galaxies are not representative of the
entire population of galaxies producing damped Ly$\alpha$ absorptions
because the masking effect makes difficult to identify faint and small
galaxies as DLA galaxies.

To evaluate the masking effect on the luminosity more precisely, we vary
the limit of angular size $\theta_{\rm th}$.  Figure 7 shows the results
for $\theta_{\rm th}=1,2$ and 3 arcsec, respectively.  The large
$\theta_{\rm th}$ produces better agreement with the luminous DLA
galaxies.  Because the large $\theta_{\rm th}$ causes faint galaxies not
to be identified as DLA galaxies statistically, our results show that
the average luminosity increases and matches better with the
observations.

We have shown that the masking effect should be taken seriously to the
identification of DLA galaxies at high redshift although the limits
depend on quasar fields. Our model also predicts that the missing rate
of DLA galaxies by the masking effect attains 60-90 \% in observations
if the observable angular size is as small as 1 arcsec and if compact
galaxies, $\la 3$ kpc, contribute significantly to the population of DLA
systems.

\section{Predicted Properties \& Implications}

\subsection{Some Properties Expected by Radio Observations}

Here, we explore another way to studying DLA galaxies to avoid the
selection biases.  Are there any methods to identify galaxies producing
damped Ly$\alpha$ absorptions apart from the masking effect?  It is
expected that radio surveys can detect this kind of galaxies at low
redshift if they have \ion{H}{1} column densities as large as those
measured in DLA systems.  We focus on the \ion{H}{1} mass in DLA systems
which correlates with observational properties in radio observations.
The cross section of radio sources also provides a useful clue to
studying the distribution of \ion{H}{1} gas in the gaseous disk. First,
a relation between \ion{H}{1} masses and sizes in DLA systems is plotted
as a solid line with $1 \sigma$ error bars in Figure 8(a).  We find that
the logarithmic cross section, $\log\sigma$, is linearly proportional to
the logarithmic \ion{H}{1} mass, $\log M_{\rm HI}$, i.e.,$\sigma \propto
M_{\rm HI}^{\alpha}$.  The relation is fitted by averaged least-squares
and we find that the slope $\alpha$ is $0.97 \pm 0.01$ in the range of
\ion{H}{1} mass $10^{6.5} \le M_{\rm HI}/M_{\odot} \le 10^{10.5}$.

Recently, \citet{RS03} observed local \ion{H}{1}-rich galaxies
identified by a blind $21$ cm survey: the Arecibo Dual-Beam Survey
(ADBS).  The ADBS sample consists of approximately 260 galaxies.  They
focused on the \ion{H}{1}-selected galaxies which have \ion{H}{1} column
densities comparable to those in DLA systems.  If a bright quasar
existed behind them, such \ion{H}{1}-selected galaxies should give arise
to damped Ly$\alpha$ absorption in the quasar spectrum.  Including other
data provided by different observations, they found a correlation
between  the cross section and the \ion{H}{1} mass such as $\log \sigma
\propto \log M_{\rm HI}$.  The range of the observational data spreads
within the dashed box region presented in Figure 8(a), which spans a
wide range of masses $10^{6.5} \la M_{\rm HI}/M_{\odot} \la 10^{10.5}$.
The slope of the relation is estimated as $1.004 \pm 0.021$ in the
sample.  We find that this slope is entirely consistent with our
results.  The predicted cross-sections are also within the range of the
observational data.  One might consider that the observed cross sections
are somewhat larger than our results.  The blind $21$ cm emission-line
survey has typical resolutions $\sim 10-60$ arcsec which are lower than
those in photometric surveys.  Such low resolutions may systematically
lead to overestimates in the sizes of small galaxies.

We also show the averaged cross-section of DLA systems at $0 \le z \le
1$ as a dotted line.  The slope is almost identical with that of the
relation at $z=0$ (solid line) because both the cross sections and the
\ion{H}{1} masses are almost constant against redshift [see Figure 1(c)
and Figure 7 in Paper I].  We thus predict that the relation at high
redshifts would be almost identical to that at $z=0$.  Figure 8(b) shows
the number fraction in \ion{H}{1} mass.  As already discussed in Paper
I, we find that the average of \ion{H}{1} mass is about $10^{9}$
M$_{\odot}$ at $0 \le z \le 1$ in our model.  This result agrees the
observational feature that DLA galaxies are dominated by those with
\ion{H}{1} masses near $10^{9}$ M$_{\odot}$ \citep{RS03}.

We put emphasis on the facts, first, that {\it the tight relation
between the \ion{H}{1} mass and the cross section is found in our model
similar to the radio observations}, and second, that {\it our results
are also consistent with the properties obtained by the radio
observations in the whole observed range of \ion{H}{1} mass}.  We will
extensively discuss the properties of the \ion{H}{1}-selected galaxies
in a separate paper.

\subsection{Star Formation Rates}

Star formation rates (SFRs) provide an important clue to understanding
DLA galaxies.  Figures 9 present the number fractions as a function of
both SFR and SFR per unit area in DLA galaxies at $z = 0$ and $1$,
respectively.  The SFR per unit area is simply estimated as SFR divided
by the disk area.  All galaxies are simply assumed to be face-on to us
when computing the area because the effects of the inclination are
negligible (see Figure 4 in Paper I).  We here discard spheroidal
galaxies because they have much less amount of \ion{H}{1} gas than
spirals and then much lower SFRs.  This assumption is suitable for
calculating the SFR per unit area.

In Figure 9 (a), we find that DLA galaxies have SFRs which widely ranges
from $10^{-6}$ to $10^{2}$ M$_{\odot}$ yr$^{-1}$.  At $z=1$, the
distribution of SFRs has a large variance and a long tail toward low
SFRs $\la 10^{-2}$ M$_{\odot}$ yr$^{-1}$, while the variance at $z=0$ is
smaller than that at $z=1$. Because less massive galaxies become
dominant at high redshift (Figure 8 in Paper I) and they have long
timescales of star formation, galaxies with low SFRs increases toward
high redshift. We also find that the mean SFRs are $\sim 10^{-2}$
M$_{\odot}$ yr$^{-1}$ at $0 \le z \le 1$. In Figure 9 (b), the
distributions of SFRs per unit area at $z=0$ and $1$ are quite similar
to each other with the mean values $\sim 10^{-2.8}$ M$_{\odot}$
yr$^{-1}$ kpc$^{-2}$.

Figure 10 shows various distributions of SFRs. We plot the
correlations of SFRs with redshifts, luminosities, \ion{H}{1} column
densities, sizes and surface brightnesses.  Figure 10(a) shows the
redshift evolution of SFRs. This indicates that the SFRs of DLA systems
are $\sim 10^{-2}$ M$_{\odot}$ yr$^{-1}$ at $z=0-1$ and the evolution is
quite moderate similarly to the stellar mass evolution (see Figure 7 in
Paper I).  It appears that the variances of SFRs decrease toward low
redshift as also shown in Figure 9(a).  Figures 10(b) depicts the
relation between SFRs and luminosities.  Clearly, brighter galaxies have
higher SFRs. For example, DLA galaxies brighter than $L/L^{*} \sim 0.5$
have SFRs larger than $\sim 10$ M$_{\odot}$ yr$^{-1}$ while the mean SFR
is much lower, $\sim 10^{-2}$ M$_{\odot}$ yr$^{-1}$.  It is also
noticeable that DLA systems with high $N_{\rm HI}$, large $b$ and high
(small value of) $\mu$ have larger SFRs in Figures 10(c),(d) and (e).

Consequently, our results imply that DLA galaxies have wide-ranged
values of SFRs.  In observational studies, SFRs in various types of
galaxies have been derived and discussed by various authors.  
Measurements of SFRs give the most likely value of $\sim 3$ M$_{\odot}$
yr$^{-1}$ for our Galaxy \citep{C00}. \citet{P97} reported that SFRs of
local blue compact galaxies are $\sim 0.3-0.5$ M$_{\odot}$ yr$^{-1}$.
SFRs of LSB galaxies estimated by \citet{vH00} are about $0.03-0.2$
M$_{\odot}$ yr$^{-1}$, using an empirical relation between H$\alpha$ and
$I$-band surface brightnesses in galactic disks.  For dwarf and LSB
galaxies, \citet{vZ01} compiled their observational characteristics and
estimated the SFRs using optical images in $UBV$ and H$\alpha$
passbands. The SFRs widely span at least 4 orders of magnitude, $\sim
10^{-5}-10^{-1}$ M$_{\odot}$ yr$^{-1}$.  The wide-ranged SFRs are
broadly consistent with those predicted by our model.  In this sample,
the galaxies have central surface brightnesses between $20.5$ and $25$
mag arcsec$^{-1}$ and scale lengths of the disk from about $0.2$ to
$4.3$ kpc. These properties are quite similar to those of DLA systems at
$z \sim 0$ in our model. \citet{BTJ01} found a nearby DLA galaxy
(SBS1543+593) at $z=0.009$ from photometric images obtained by a
ground-based telescope and {\it Hubble Space Telescope (HST)}.  Because
this DLA system is the closest to us, it is expected to provide some
clues to revealing the various properties of DLA galaxies.  Recently,
\citet{S04} reported a more detailed study of this galaxy and found that
this has typical properties of LSB galaxies: the central surface
brightness $\mu_{\rm B}(0)=22.8 \pm 0.3$ mag arcsec$^{-2}$ and the SFR
$\sim 0.006$ M$_{\odot}$ yr$^{-1}$.  These properties also agree with
those in our model. Together with these observational facts, our result
leads to a conclusion that DLA systems mainly comprise dwarf galaxies in
which the SFRs are low comparable to those in LSB galaxies.

If DLA galaxies are faint and compact, it happens that most of the
stellar counterparts cannot be detected by selection effects, which
would prevent from estimating the SFRs accurately by using emission
lines.  Absorption lines potentially provide us useful information on
SFRs even if DLA galaxies are faint and compact.  Recently, Wolfe,
Prochaska, \& Gawiser (2003a,b) successfully estimated the SFRs per unit
area of about 30 galaxies at $z\ga 2$ by a new method using
\ion{C}{2}$^{*}$ $\lambda~1335.7$ absorption line.  They inferred the
SFRs per unit area from the heating rate by far-ultraviolet radiation
from massive stars under the assumption that the heating rate is equal
to the rate of cooling by \ion{C}{2}$^{*}$ line emission.  Generally,
the cold gas consists of two-phase media comprising a cold neutral
medium (CNM) and a warm neutral medium (WNM). These media can stably
exist in pressure equilibrium under some conditions which are determined
by a thermal balance of the two phases \citep[e.g.][]{MO77, W95}.  For
example, \citet{W95} investigated the stable condition of the CNM and
the WNM in the interstellar medium (ISM). They found that neutral gas
exists in two stable phases, the CNM with typical temperature $T \sim
10^{2}$ K and a typical number density $n \sim 0.1$ cm$^{-3}$, and the
WNM with $T \sim 8000$ K and $n \sim 10$ cm$^{-3}$. \citet{W03a} showed
stable conditions of the CNM and the WNM in DLA systems similar to those
in the ISM of the Milky Way. In our model at present stage, we assume
that disk gas comprises uniformly one-phase medium for simplicity. When
the disk gas is uniformly distributed throughout plane-parallel disks
with radius $r \sim 3$ kpc and scale-height $h \sim 0.1$ kpc, the
typical number density is $n \sim 0.1$($N_{\rm HI}/10^{20.5}$
cm$^{-2}$)($h/0.1$ kpc)$^{-1}$ cm$^{-3}$.  In a low-density medium ($n
\la 0.1$ cm$^{-3}$), cosmic rays are a dominant source of heating and
ionization \citep{W03a}. Assuming that the ionization rate by cosmic
rays is proportional to the star formation rate, the ionization fraction
depends on the star formation rate in addition to the number density and
temperature. The physical conditions typically predicted by our model,
$N_{\rm HI} \sim 10^{20.5}$ cm$^{-2}$, the SFR per unit area$\sim
10^{-2.8}$ M$_{\odot}$ yr$^{-1}$ kpc$^{-2}$, and $n \sim 0.1$ cm$^{-3}$,
would result in the ionization fraction $x$($= n_{\rm e}/n$) less than
$0.1$ \citep[]{W95, W03a}, and suggests that the disk gas in our
calculation roughly corresponds to the WNM for two-phase
model. \citet{W03a} estimated the SFRs $\sim 10^{-3}-10^{-2}$
M$_{\odot}$ yr$^{-1}$ kpc$^{-2}$ for DLA systems in case that
\ion{C}{2}$^{*}$ absorption occurs in the CNM, and $\sim
10^{-2}-10^{-1}$ M$_{\odot}$ yr$^{-1}$ kpc$^{-2}$ for DLA systems in
case that \ion{C}{2}$^{*}$ absorption occurs in the WNM. Interestingly,
the former SFRs agree well with those in our model.  This might imply
that \ion{C}{2}$^{*}$ absorption occurs in the CNM at $z \ga 2$ if the
evolution of SFRs are moderate against redshift. It is also possible
that the SFRs decrease toward low redshift ($z < 2$) in the WNM in which
\ion{C}{2}$^{*}$ absorption arises.  Indeed, since we assume that the
gas is uniformly distributed in the disk, our modeling for disk gas has
not reached a crucial stage to compare our results with the estimated
SFRs precisely. It would be valuable to investigate the evolution of
multi-phase structure in DLA systems in a subsequent study. Recent
hydrodynamical simulations have estimated SFRs of DLA systems. For
example, using SPH simulation, \citet{NSH04} found that simulated DLA
systems have SFRs per unit area as low as $\sim 10^{-3}-10^{-2}$
M$_{\odot}$ yr$^{-1}$ kpc$^{-2}$.  Their results should be suggestive to
recognize that the simulated SFRs are consistent with our results.  We
should note that, however, their simulation has several limitations such
as numerical resolutions, the narrow range of parameters related with
star formation and supernova feedback, and so on, and also that they
predicted about an order of magnitude higher metallicities of DLA
systems than observed values.

Figure 11 shows the SFR per unit area as a function of the \ion{H}{1}
column density at redshifts $z=0$ and $1$ in comparison with the
Kennicutt law, which is an empirical relation between the SFR surface
density and the \ion{H}{1} column density of local galaxies provided by
\citet{K98},\\

SFR surface density
\begin{eqnarray}
=(2.5 \pm 0.7) \times 10^{-4} 
\left (
\frac{\displaystyle N_{\rm HI}}{\displaystyle 1.25 \times 10^{20} 
{\rm cm^{-2}}}
\right )^{1.4 \pm 0.15} \nonumber \\
{\rm M_{\odot} yr^{-1} kpc^{-2}}. 
\end{eqnarray}
This relation is also represented in Figure 11 by the shaded region.  We
find that our SFRs broadly follow the Kennicutt law. In other studies,
this trend is also emerged at redshift $z \ga 2$ (Wolfe, Prochaska \&
Gawiser 2003a,b). In a numerical simulation, \citet{NSH04}, furthermore,
confirmed this tendency which seems to be independent of the strength of
galactic winds. In addition with our results, these suggest that the
SFRs follow the Kennicutt law at both low and high redshifts. This
implies that the evolution of SFRs in DLA systems should be very
moderate against redshift. Including other fundamental properties such
as luminosities and sizes at high redshift, the SFR evolution at $z > 1$
will be also presented and discussed in a subsequent paper.

\subsection{Sizes and \ion{H}{1} column density distributions of DLA galaxies}

Finally, we examine how the masking effect changes other properties of
DLA galaxies. First, Figure 12 shows the radial size $b$ as a
function of $B$-band absolute magnitude brighter than $M_{\rm B} \sim
-17$ mag.  Our result shows a trend that bright DLA galaxies are likely
to be observed with large impact parameters.  The mean radial size is
apparently larger than the observational data because the radial size
provides an upper limit of impact parameters.  This scaling relation is
fitted by averaged least-squares, $b \propto M_{\rm B}^{\alpha}$, and we
find the slope $\alpha = -0.18 \pm 0.02$ in the range of
magnitude $-21 \le M_{\rm B} \le -16.5$ mag. Taking into account the
masking effect, we also show the $b-M_{\rm B}$ relation as the dashed
line in this figure.  It is also evident that the masking effect with
$\theta_{\rm th}=1$ arcsec hardly changes the slope for bright galaxies
because the masking effect reduces the number of only compact and faint
galaxies.  Recently, \citet{CL03} observationally found a scaling
relation between the \ion{H}{1}-disk size $R$ and the $B$-band
luminosity $L_{\rm B}$ of DLA galaxies, $R \propto L_{\rm B}^{\beta}$
with $\beta=0.26^{+0.24}_{-0.06}$.  This relation is also plotted as the
dotted line in Figure 12. We find that the scaling relation is
consistent with our predicted one within 1$\sigma$ scatters. For
example, our relation has the indices $\beta=0.35 \pm 0.02$ and $0.45
\pm 0.01$ with and without the masking effect, respectively. Thus the
tight relationship between the radial size and the magnitude found in
our model, which suggests that brighter galaxies are larger, is
consistent with the observations.

Second, we calculate a neutral hydrogen column-density distribution of
{\it DLA galaxies} taking into account the masking
effect.  Figure 13 shows the column density distribution at $z=1$.  The
data points are taken from \citet{SW00} \footnote{We adjust the data
points for the cosmological model (LCDM) adopted here.}.  We also take
additional data reported by \citet{RT00} which include a sample at low
redshift $ \langle z \rangle \sim 0.8$.  In Paper I (Figures 3 and 5),
we reproduced the column density distribution of DLA systems especially for
low-redshift ones.  In Figure 13, the solid and dashed lines indicate
the distributions of DLA galaxies with and without the masking effect,
respectively.  The masking effect is the most significant at $z=1$
because an angular size of a constant physical size approaches a minimum
around $z=1$.  Therefore, this effect reduces the number of DLA galaxies
over the whole range of column density.  Even if $\theta_{\rm th}=1$
arcsec, we find that the column density distributions can be still
consistent with the observational data (dashed line in Figure 13).  We
thus predict that optically-selected DLA galaxies also have the column
density distribution similar to that of all Damped Ly$\alpha$ absorbers 
even if the selection effect exists.

\section{Conclusions and Discussion}

We have investigated damped Ly$\alpha$ absorbing galaxies at redshifts
$0 \le z \le 1$ in the hierarchical structure formation scenario using a
semi-analytic galaxy formation model.  In the previous study
\citep{ONGY}, we found that our model can reproduce many fundamental
properties of DLA systems such as the metallicity evolution, the column
density distribution and the mass density of cold gas in addition to
those of local galaxies \citep{NTGY}.  In this paper, we have focused
our attention on their host galaxies that give arise to damped
Ly$\alpha$ absorption at $0 \le z \le 1$ and use the model to calculate
the observable properties.  The main conclusions are the followings:

\begin{enumerate}
 \item Most of DLA galaxies producing damped Ly$\alpha$ absorption lines
 in quasar spectra are faint and compact. Their typical size is $\sim 3$
 kpc, and the mean surface brightnesses are $\sim 22$ mag arcsec$^{-2}$
 at $z \sim 0$ and $\sim 27$ mag arcsec$^{-2}$ at $z \sim 1$,
 respectively.  Some selection biases are required for those to have
 fundamental properties consistent with those of DLA galaxies observed
 in optical and near-infrared images \citep{R03}.

\item Two selection biases were studied here.  First is a bias caused by
low surface-brightness galaxies.  A typical limit of surface brightness
in observations, $\mu_{\rm th}=25$ mag arcsec$^{-2}$, has a negligible
effect.  Second is {\it the masking effect}, under which only large DLA
galaxies are detectable because small ones must reside in close
proximity to a quasar line of sight where the quasar PSF dominates.
Considering a typical masking angular size $\theta_{\rm th}=1$ arcsec,
this effect is significant and makes the distributions of fundamental
properties of DLA galaxies much better agree with the observations.

\item The missing rate of DLA galaxies by the masking effect attains
$60-90~\%$ if low-luminosity galaxies with small impact parameters
($\sim 3$ kpc) significantly contribute to the population of DLA
systems.

\item A tight relation between the \ion{H}{1} mass and the cross section
was confirmed in DLA systems. We also found that \ion{H}{1}-rich
galaxies with $10^{9}$ M$_\odot$ mainly contribute to the population of
DLA systems at $z \sim 0$.  These results are entirely consistent with
the properties of \ion{H}{1}-selected galaxies in a radio survey
\citep[{\it the Arecibo Dual-Beam Survey},][]{RS03}.  The investigations
by such blind radio surveys could provide alternative possibilities for
exploring DLA galaxies apart from the selection biases in photometric
surveys.

\item DLA galaxies display a wide range in SFRs with the mean about
$10^{-2}$ M$_{\odot}$ yr$^{-1}$. This suggests that DLA galaxies consist
of dwarf galaxies in which SFRs are low comparable to those in LSB
galaxies.\\
\end{enumerate}

Although more data would be required to confirm these conclusions, this
study suggests that {\it LSB dwarf galaxies primarily contribute to the
population of DLA systems, while massive spiral galaxies, which is less
abundant, also should be DLA systems}.

Some previous attempts have been made to detect DLA galaxies much closer
to background quasars to unveil the nature of DLA systems.  \citet{S97}
studied a DLA system at $z=0.656$ from photometric images and
spectroscopy of the quasar 3C336 field.  They concluded that there is no
galaxy brighter than $0.05 L^{*}_{\rm K}$ within $0.5$ arcsec,
corresponding to a radius $\sim 2 h^{-1}$ kpc, to the quasar line of
sight.  \citet{B01} tried to detect H$\alpha$ emission in the region
behind the PSF of the same quasar 3C336.  But they failed to detect any
H$\alpha$ emitters in the vicinity of the line of sight of the quasar
with the range from $0.24$ to $30$ h$^{-1}$ kpc.  In this observation,
they reported that the $3~\sigma$ flux limit was $\sim 3 \times
10^{-17}$ h$^{-2}$ ergs s$^{-1}$ cm$^{-2}$ for an unresolved source.
This corresponds to an SFR of $0.3$ h$^{-2}$ M$_{\odot}$ yr$^{-1}$.  Our
results show that the mean SFR at $z \sim 0.65$ is $\sim 10^{-2}$
M$_{\odot}$ yr$^{-1}$ which are much lower than the SFR corresponding to
the flux limit.  Therefore, if LSB galaxies dominate the population of
DLA systems, the low surface brightness, corresponding to the low SFR,
may prevent us from detecting H$\alpha$ emission.

If a PSF size $\theta_{\rm th}$ is constant at $z > 1$, the physical
size corresponding to $\theta_{\rm th}$ becomes smaller, so that the
masking effect would be less serious than the samples at redshifts $z
\la 1$.  \citet{K00} reported that they were able to detect an H$\alpha$
emission feature from a DLA system at $z=1.892$ at a projected
separation of $0.25$ arcsec from a line of sight toward the quasar LBQS
1210+1731, using {\it HST} NICMOS.  They concluded that the size of the
H$\alpha$ emitter would be $2-3$ kpc if it is associated with the DLA
system and the feature is not PSF artifacts.  \citet{F99}, \citet{M02},
and \citet{M04} have successfully detected five DLA galxies at $2 \le z
\le 3$. They found that all Ly$\alpha$ emitters reside in very close
proximity to background-quasar lines of sight, and concluded that they
have small impact parameters about 1-3 kpc. Although these are
observational properties of high-redshift ($z > 2$) DLA galaxies, it is
suggestive to recognize the fact that their conclusions are consistent
with our results.

It may be more likely that DLA galaxies can be identified at $z \sim 0$
than high-redshift ones.  This is partly because a physical size
corresponding to a constant $\theta_{\rm}$ becomes small toward present
enough to identify compact galaxies around a quasar line of sight, and
partly because surface brightnesses of DLA galaxies are expected to be
so high as to detect their photometric images.  Like a DLA galaxy
(SBS1543+593) at $z=0.009$ \citep{S04}, DLA galaxies at $z \sim 0$ have
advantages that their detection is less affected by the masking effect
and/or their faintness than high-redshift ones because they are close to
us.  Thus we again emphasize our conclusions that the selection biases
are very important to understand the nature of DLA galaxies and to
interpret results of photometric observations.

The radio observations offer some possibilities for exploring the nature
of \ion{H}{1}-rich galaxies such as DLA systems.  Some blind $21$ cm
surveys provide interesting information to establish fundamental
properties of local \ion{H}{1}-selected galaxies such as the \ion{H}{1}
mass function, which is the distribution function of galaxies as a
function of the \ion{H}{1} mass, the relation between the \ion{H}{1}
mass and the near-IR luminosity of their counterparts, and so on (Zwaan
et al.2003; Rosenberg \& Schneider 2003).  For example, the \ion{H}{1}
Parkes All-sky Survey (HIPASS) is an ongoing blind survey which has
samples of 1000 galaxies with the \ion{H}{1} masses $10^{6.8} \la$
M$_{\rm HI}$ $\la 10^{10.6}$ M$_{\odot}$ \citep{Z03}.  These samples
indicate that the \ion{H}{1}-selected galaxies exhibit interesting
properties such as the \ion{H}{1} mass function.  Since the HIPASS
samples consist of \ion{H}{1}-gas systems with $N_{\rm HI} \ga 10^{19}$
cm$^{-2}$ including sub-DLA systems and Lyman-limit systems, such direct
measurements of the \ion{H}{1}-gas systems can provide stringent
constraints on observations and formation theories of DLA galaxies,
sub-DLA systems and faint dwarf galaxies such as LSB galaxies.  The
studies of the optical counterparts could offer special insights to
their nature.  Moreover, their natures can be unveiled by investigating
the missing link between \ion{H}{1}-selected galaxies and optical ones.

\acknowledgments

We thank N.Gouda, S.Yoshioka, M.Enoki, H.Yahagi and T.Yano for valuable
discussions of this study and the anonymous referee for careful reading
of this manuscript and suggestions, which improved the clarity of this
presentation. K.O. thanks Arthur Wolfe for an interesting suggestion and
Kentaro Nagamine for stimulating discussions, particularly for star
formation rates in comparison with his numerical results. M.N. also
acknowledges a PPARC rolling grant for extragalactic astronomy and
cosmology and the Japan Society for the Promotion of Science for Young
Scientists (No.207).

\newpage
\begin{figure*}
\plotone{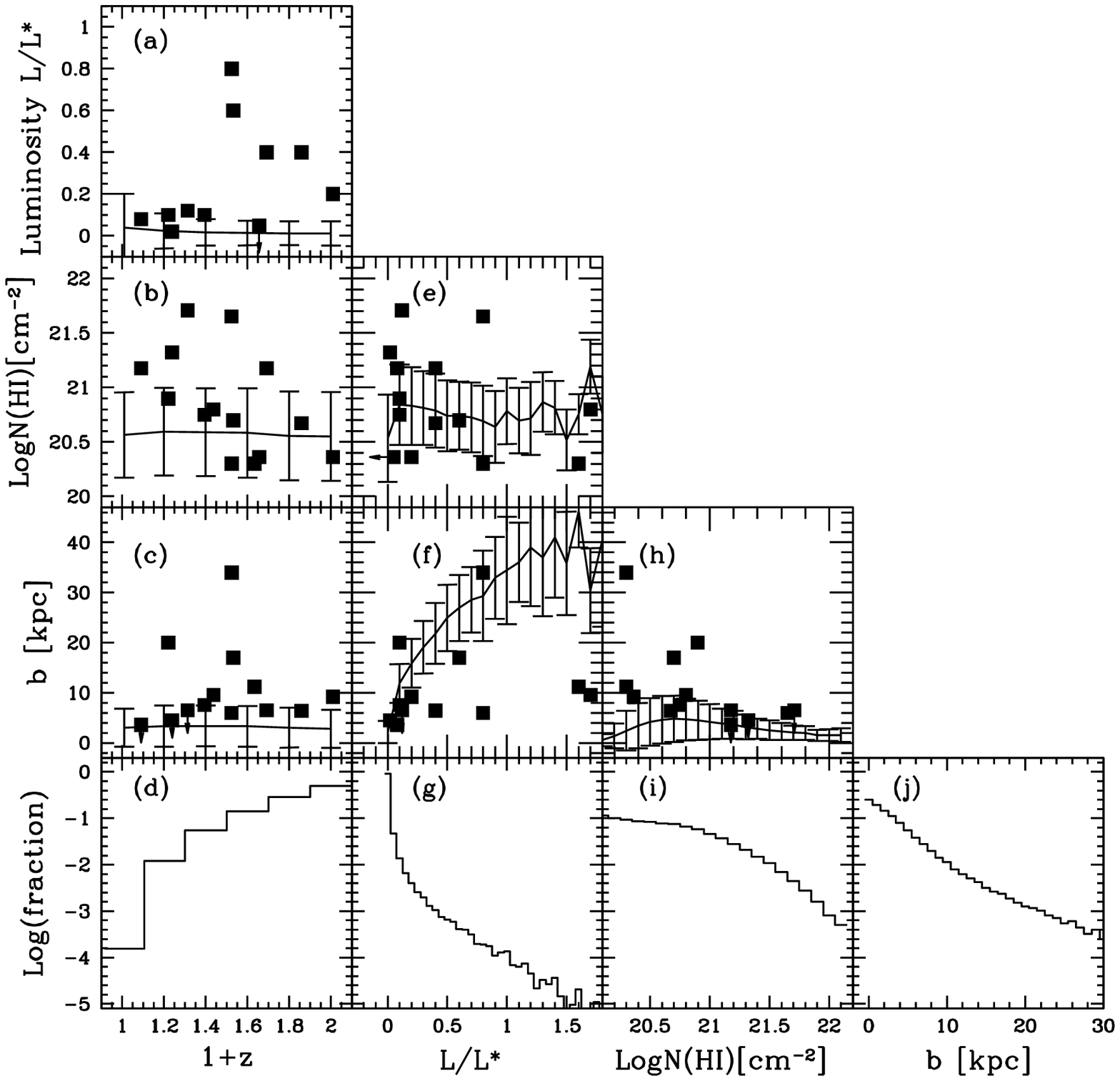}
\caption{
Fundamental properties of DLA galaxies at redshifts $0 \le z \le 1$.
Error bars with the averages indicate $1 \sigma$ errors.
The square symbols are the observational data \citep{R03}.
B-band luminosities are plotted in figures but two data for which the luminosities are
 measured in K-band only.  The data with an upper limit of the
 luminosity are the DLA system in the 3C336 quasar field. More
 information for detail is presented in \citet{R03}. 
Note that the number fraction is $\sim 0.98$ ($L \le 0.1L^{*}$) 
in panel(g).   
 } \label{fig:corrfil}
\end{figure*}

\begin{figure*}
\plotone{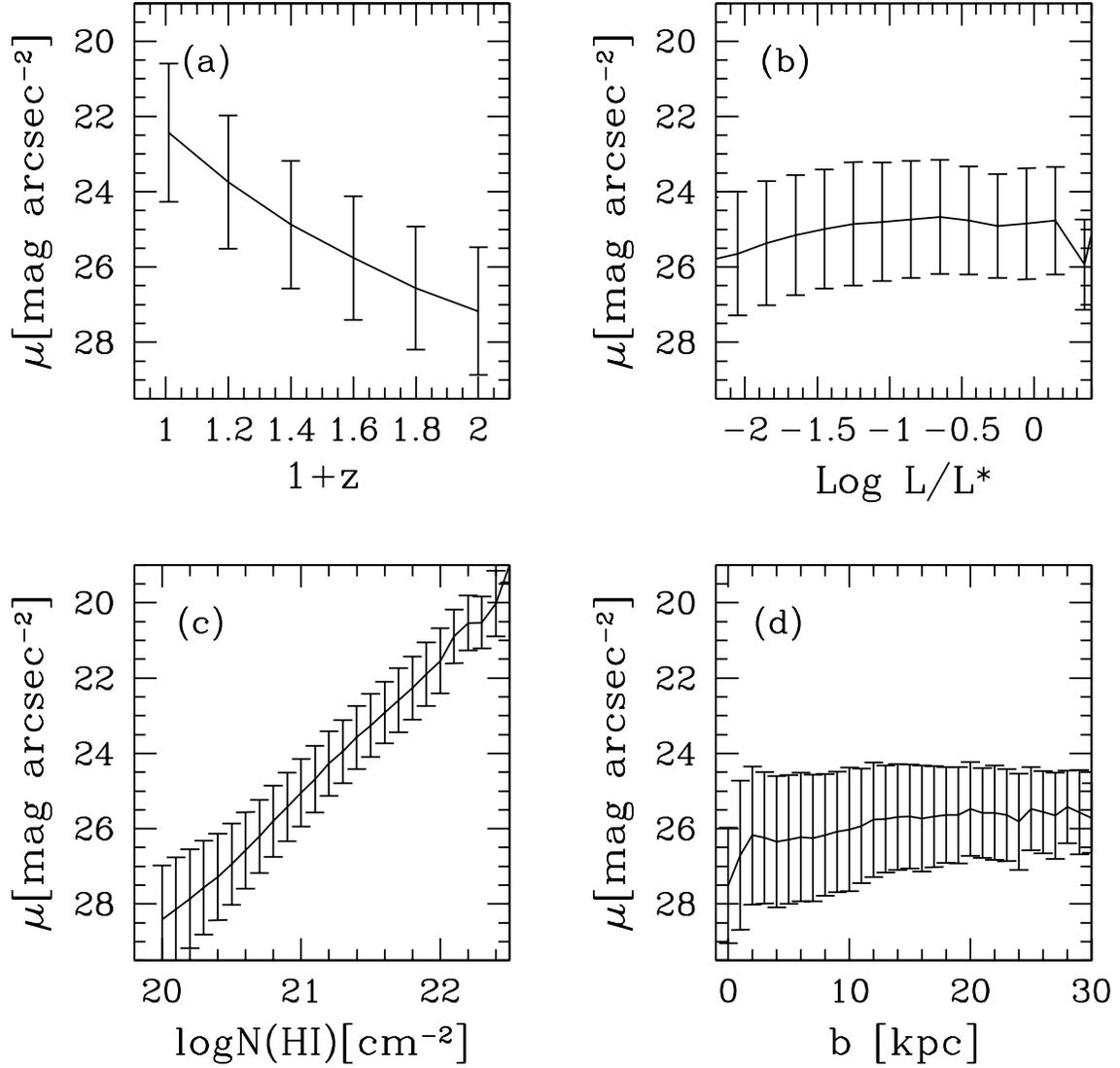}
\caption{ 
Central surfaceness brightness $\mu$ in B-band of DLA galaxies 
as a function of (a) redshift, (b) luminosity, (c) column density and 
(d) size, respectively. 
Error bars with the averages indicate $1 \sigma$ errors.
} \label{fig:corrfil}
\end{figure*}

\begin{figure*}
\plotone{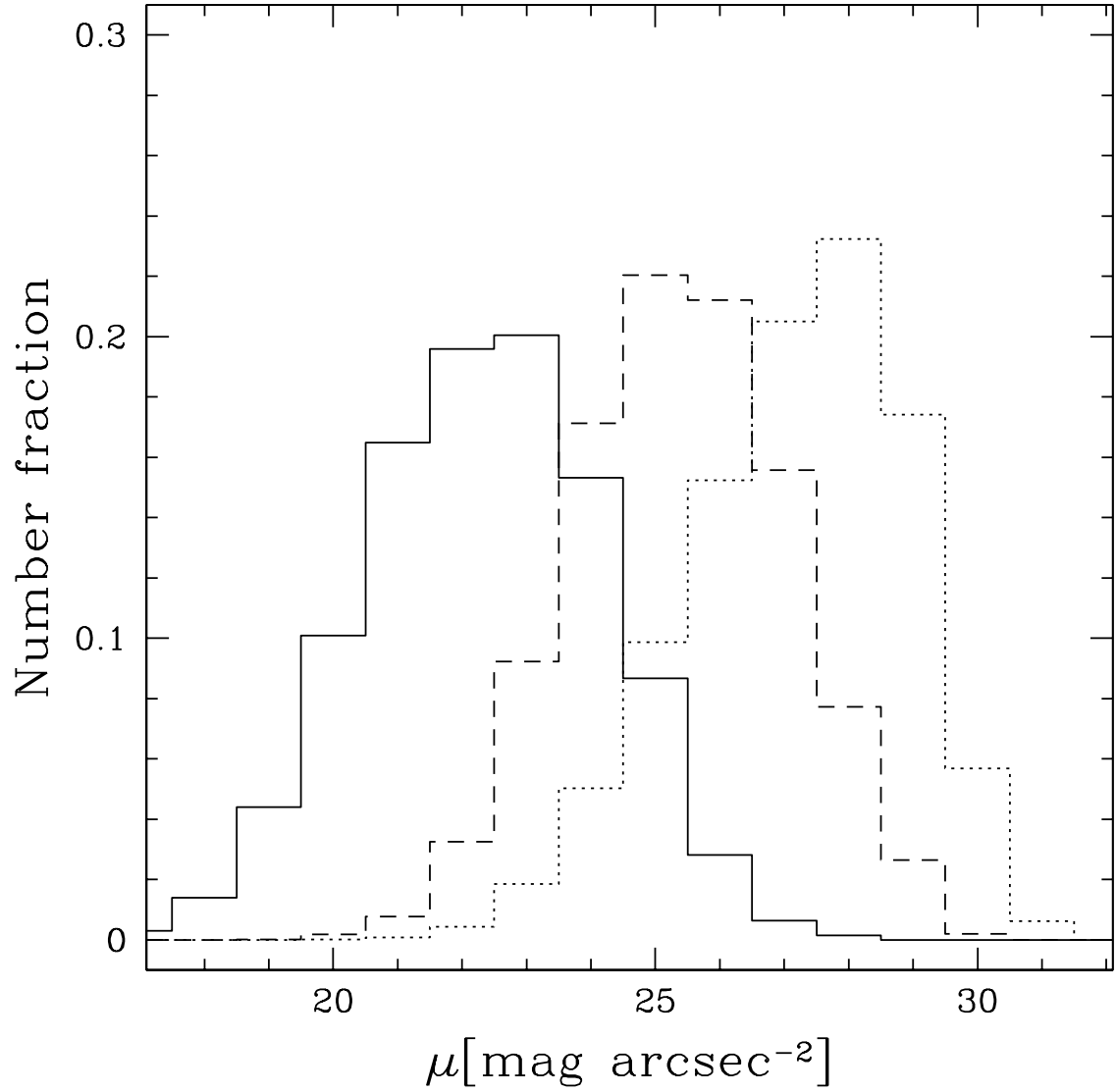}
\caption{
Number fractions of DLA galaxies as a function of surface brightness $\mu$
 in B-band at redshifts $z=0$ (solid line), $0.5$ (dashed line) and 
$1$ (dotted line). 
} \label{fig:corrfil}
\end{figure*}

\begin{figure*}
\plotone{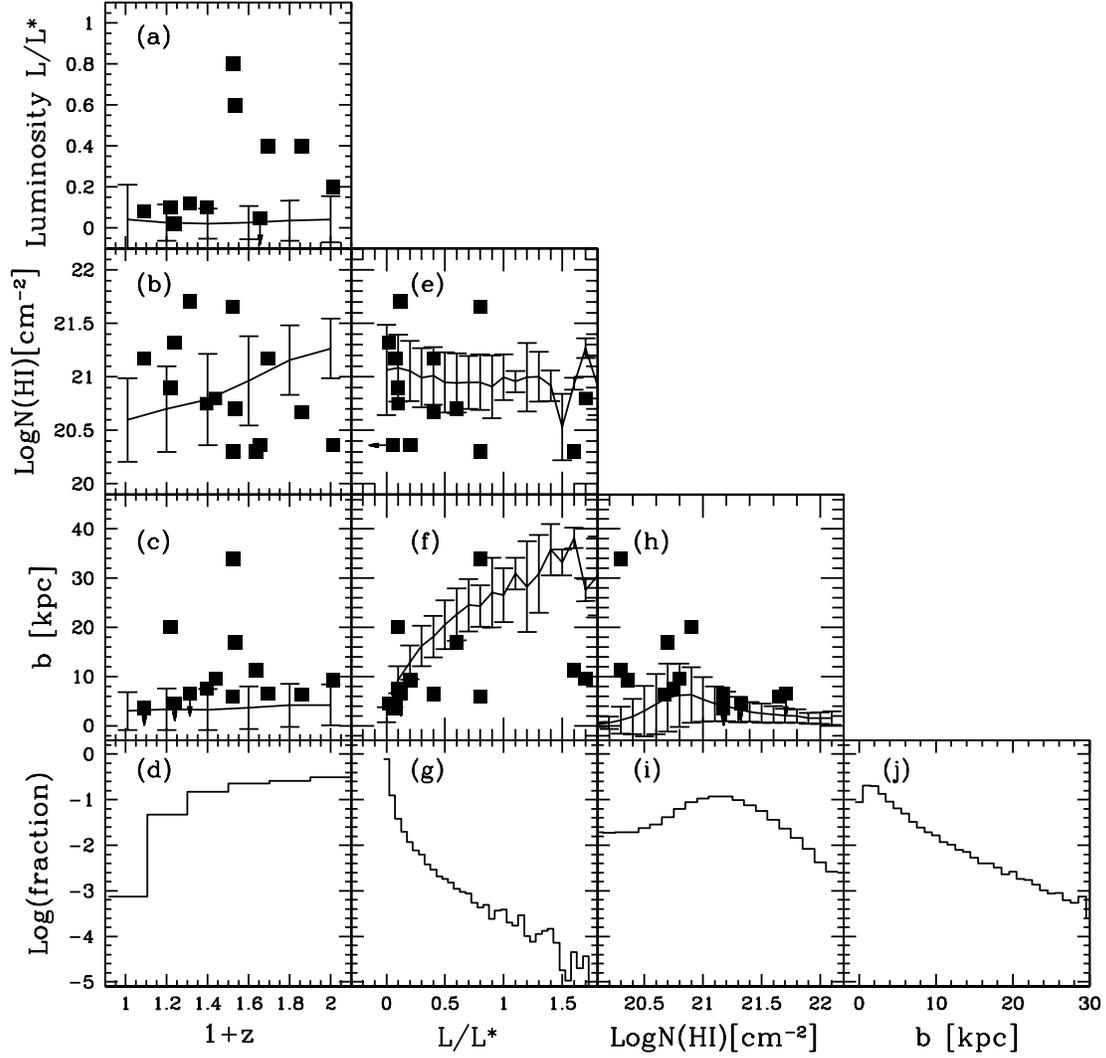}
\caption{
Fundamental properties of DLA galaxies at redshifts $0 \le z \le 1$ if 
the central  surface brightness $\mu < 25$ [mag arcsec$^{-2}$].
Error bars with the averages indicate $1 \sigma$ errors.
The square symbols are the observational data \citep{R03}.
The number fractions in panels (d),(g),(i) and (j) are defined 
as the ratios the number of DLA galaxies per
bin to that of DLA galaxies which fulfill the selection criteria.
Note that the number fraction is $\sim 0.94$ ($L \le 0.1L^{*}$) 
in panel(g).   
 } \label{fig:corrfil}
\end{figure*}

\begin{figure*}
\plotone{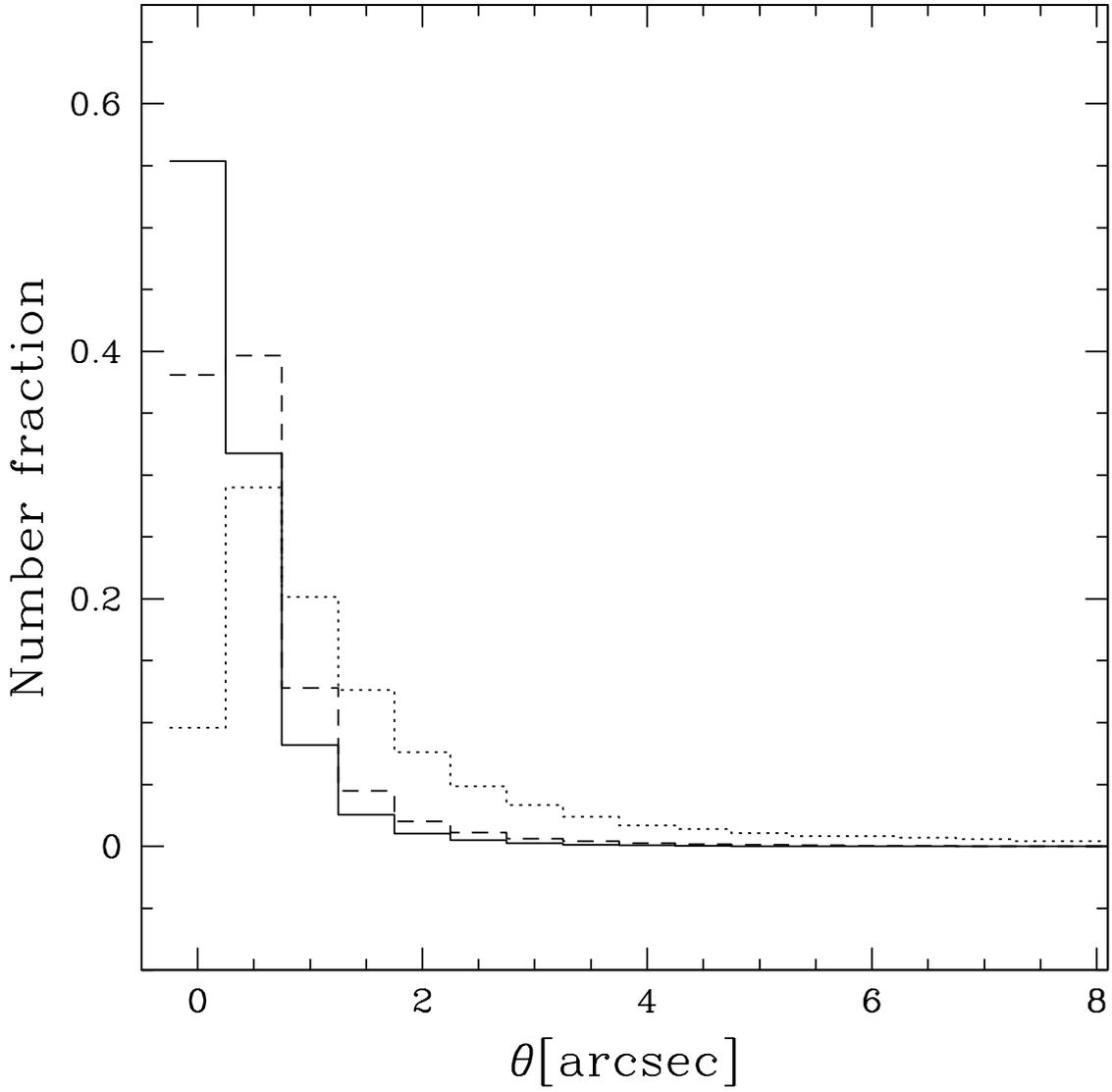}
\caption{
Number fractions of DLA galaxies as a faction of angular size
 $\theta$
 [arcsec] at $z=0.1$ (dotted line),$0.5$ (dashed line) and $1$ (solid line).  
} \label{fig:corrfil}
\end{figure*}

\begin{figure*}
\plotone{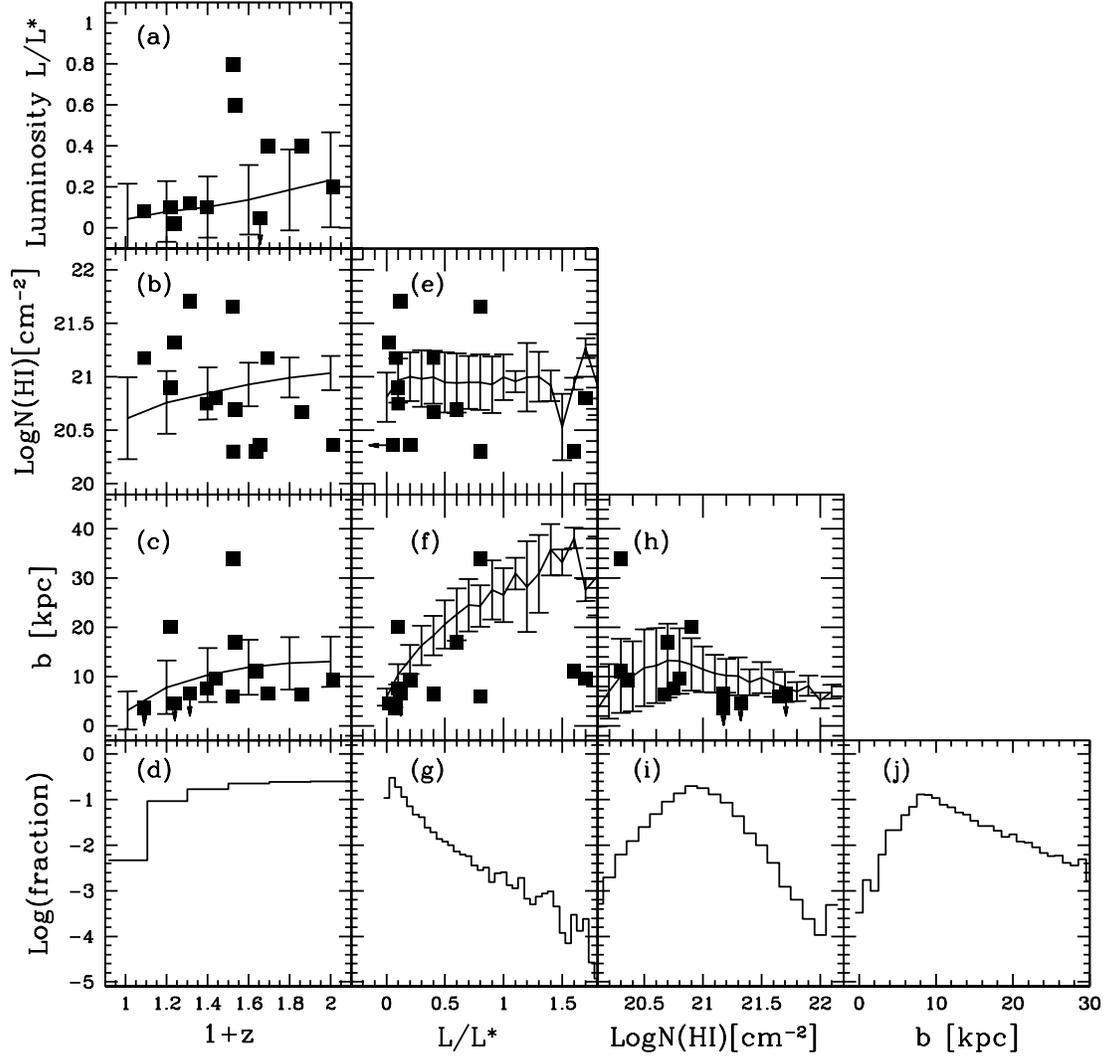}
\caption{
Fundamental properties of DLA galaxies at redshifts $0 \le z \le 1$ if
 the angular size $\theta > 1$ [arcsec] 
in addition to the surface brightness limit 
$\mu_{\rm th}=25$ [mag arcsec$^{-2}$].
Error bars with the averages indicate $1 \sigma$ errors.
The number fractions in panels (d),(g),(i) and (j) are defined 
as in Figure 4. 
The square symbols are the observational data \citep{R03}.
Note that the number fraction is $\sim 0.60$ ($L \le 0.1L^{*}$) 
in panel(g).   
 } \label{fig:corrfil}
\end{figure*}

\begin{figure*}
\plotone{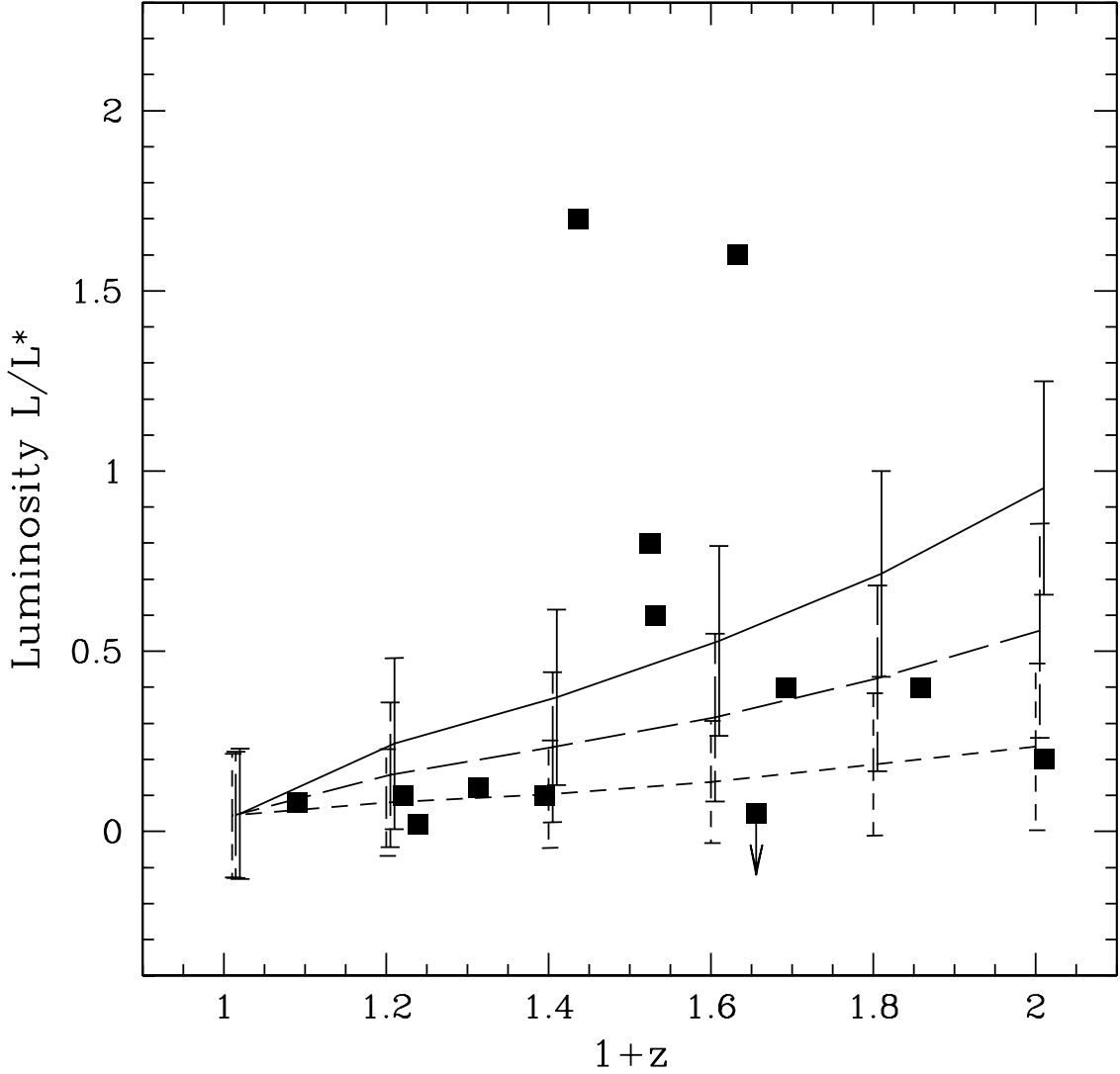}
\caption{
Luminosity evolution of DLA galaxies taking into account angular-size limitations: 
$\theta_{\rm th}=$1 (short-dashed line),2 (long-dashed line) 
and 3 (solid line)[arcsec] 
in addition to the surface brightness limit 
$\mu_{\rm th}=25$ [mag arcsec$^{-2}$]. 
Error bars with the averages indicate $1 \sigma$ errors.
The square symbols are the observational data \citep{R03}.
} \label{fig:corrfil}
\end{figure*}

\begin{figure*}
\plotone{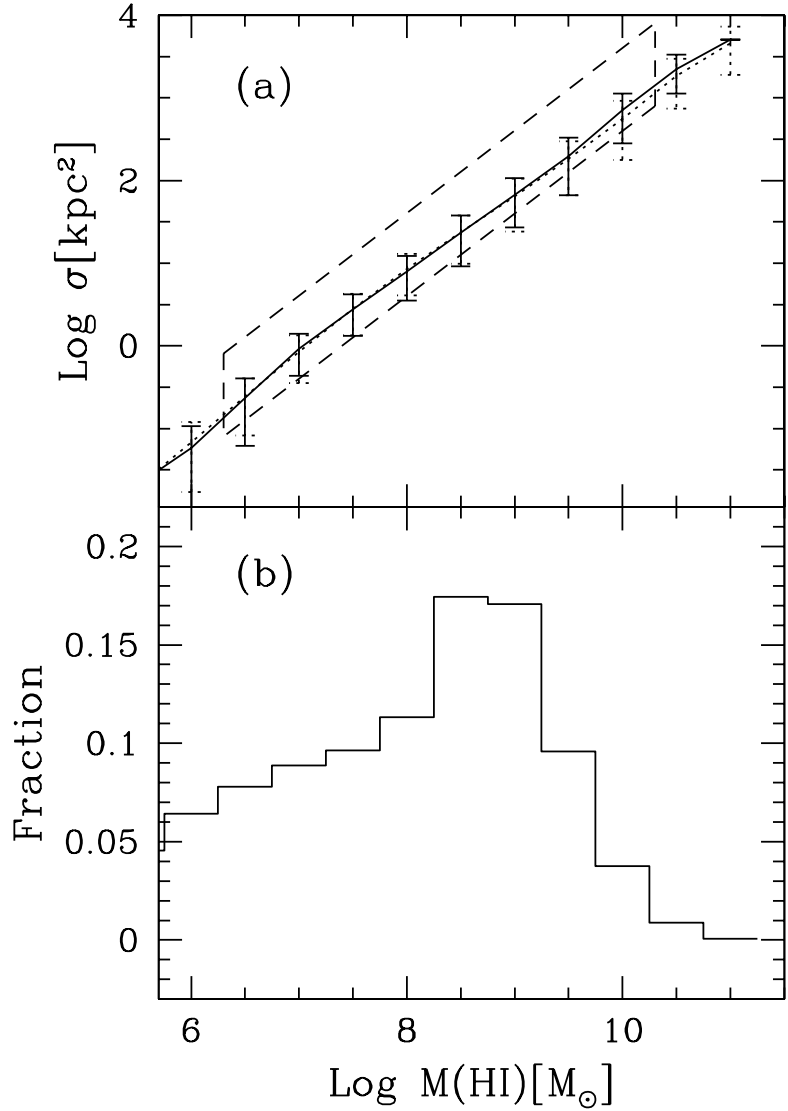}
\caption{
Relations with \ion{H}{1} mass of DLA galaxies at $0 \le z \le 1$.
(a)The cross section vs \ion{H}{1} mass. 
The average cross sections at $z=0$ are plotted as the solid line.
Error bars with the averages indicate $1 \sigma$ errors.  
The boxes enclosed by dashed lines  represent ranges of the
 observational data from blind $21$ cm surveys \citep{RS03}. 
We also show the averaged cross-section at $0 \le z \le 1$ 
as the dotted line which is almost identical with the relation at
 $z=0$ (solid line).  
(b)Number fraction of DLA galaxies as a function of \ion{H}{1} mass at 
$0 \le z \le 1$.
} \label{fig:corrfil}
\end{figure*}

\begin{figure*}
\plotone{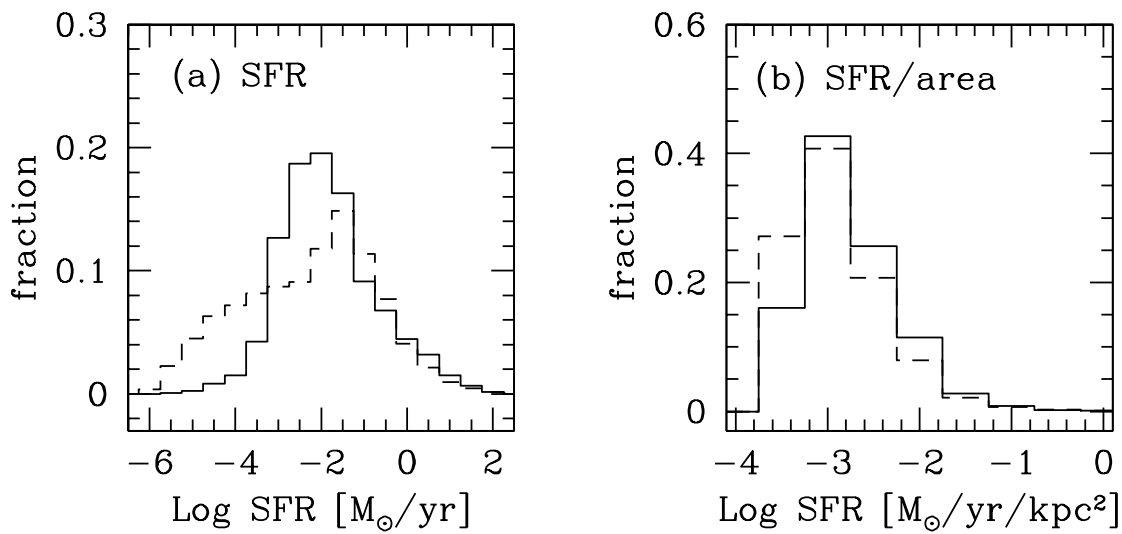}
\caption{
(a)Number fractions of DLA galaxies as a function of star formation rate 
[M$_{\odot}$ yr$^{-1}$] at $z=0$ (solid line) and 
$z=1$ (dashed line), respectively. 
(b)Number fractions as a function of star formation rate per unit area 
[M$_{\odot}$ yr$^{-1}$ kpc$^{-2}$] are also shown at $z=0$ (solid line) 
and $z=1$ (dashed line), respectively.   
} \label{fig:corrfil}
\end{figure*}

\begin{figure*}
\plotone{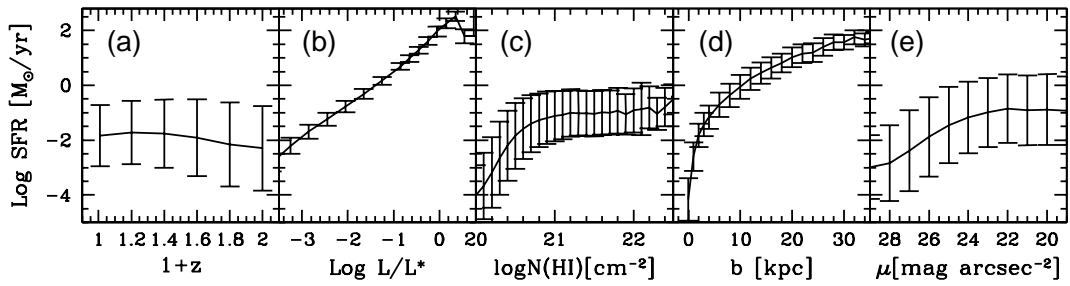}
\caption{ 
Star formation rates of DLA galaxies as a function of (a)redshift, 
(b)absolute luminosity in B-band, (c)\ion{H}{1} column density, 
(d)size and (e)surface brightness in B-band, 
respectively. 
Error bars with the averages indicate $1 \sigma$ errors.
} \label{fig:corrfil}
\end{figure*}

\begin{figure*}
\plotone{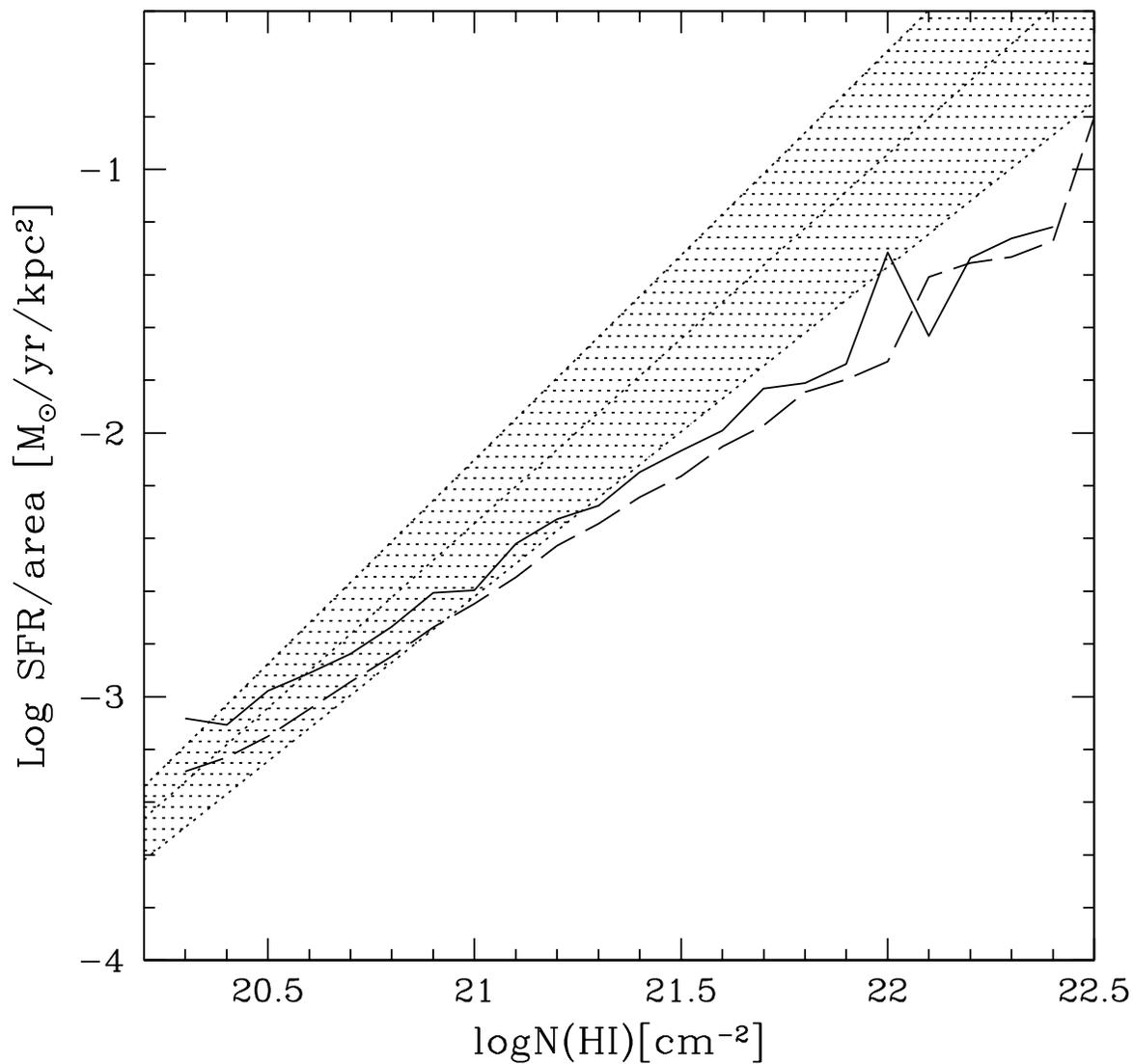}
\caption{
Star formation rates per unit area of DLA galaxies as a function of 
\ion{H}{1} column density at redshift $z=0$ (solid line) and $z=1$ (long-dashed
 line) in comparison with the Kennicutt law (1998), respectively.  
The shaded area indicates the region of SFRs given by equation(3) which 
includes systematic errors in the SFRs. 
} \label{fig:corrfil}
\end{figure*}

\begin{figure*}
\plotone{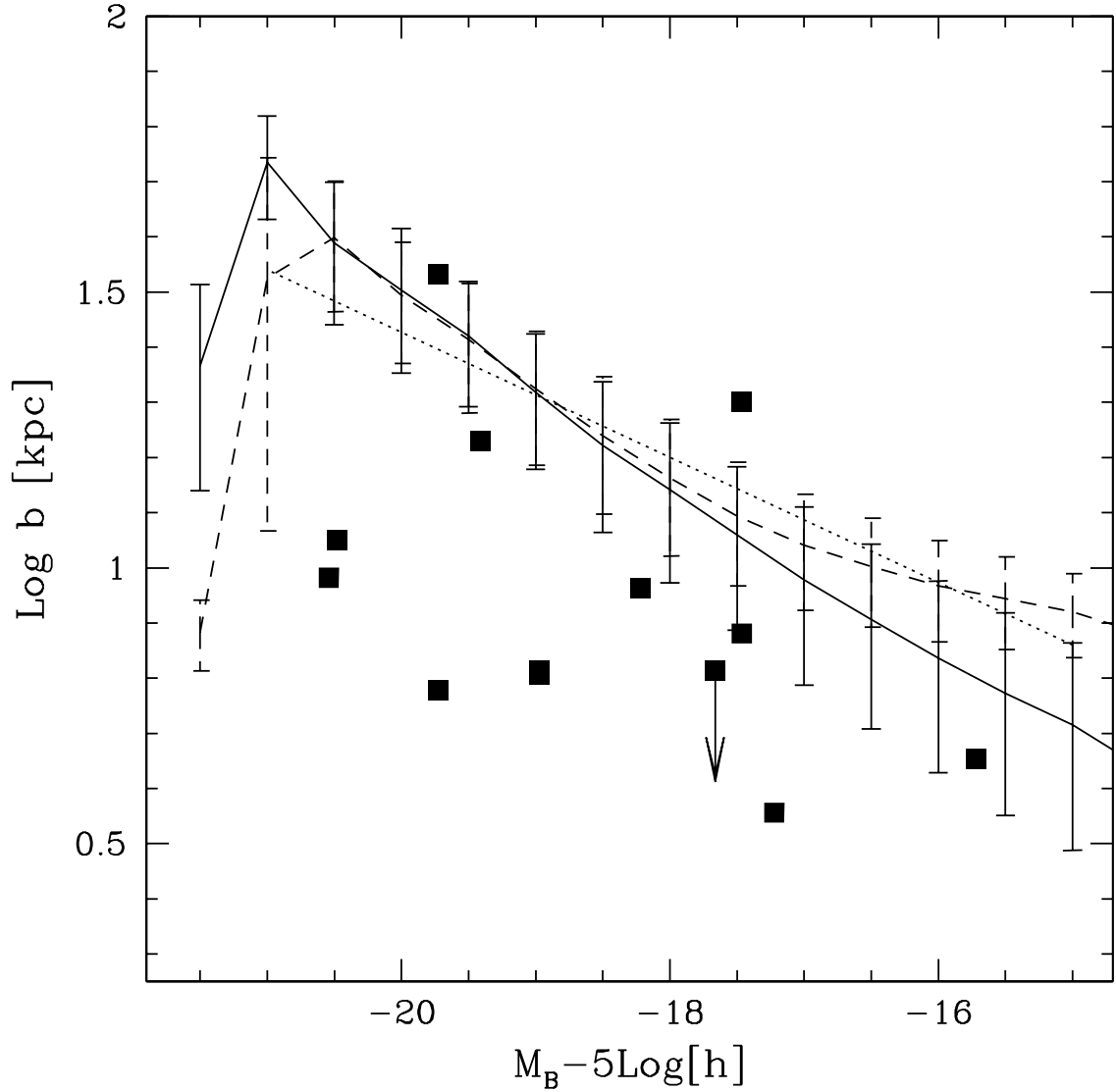}
\caption{
The radial size as a function of B-band absolute magnitude: 
no angular size limitation (solid line) and $\theta > 1$ [arcsec] (dashed line), respectively. 
As shown in Figure1, the square symbols are the observational data
 \citep{R03}.
Dotted line shows the scaling relation between magnitudes and 
\ion{H}{1} sizes provided by \citet{CL03}.
Error bars with the averages indicate $1 \sigma$ errors.
} \label{fig:corrfil}
\end{figure*}

\begin{figure*}
\plotone{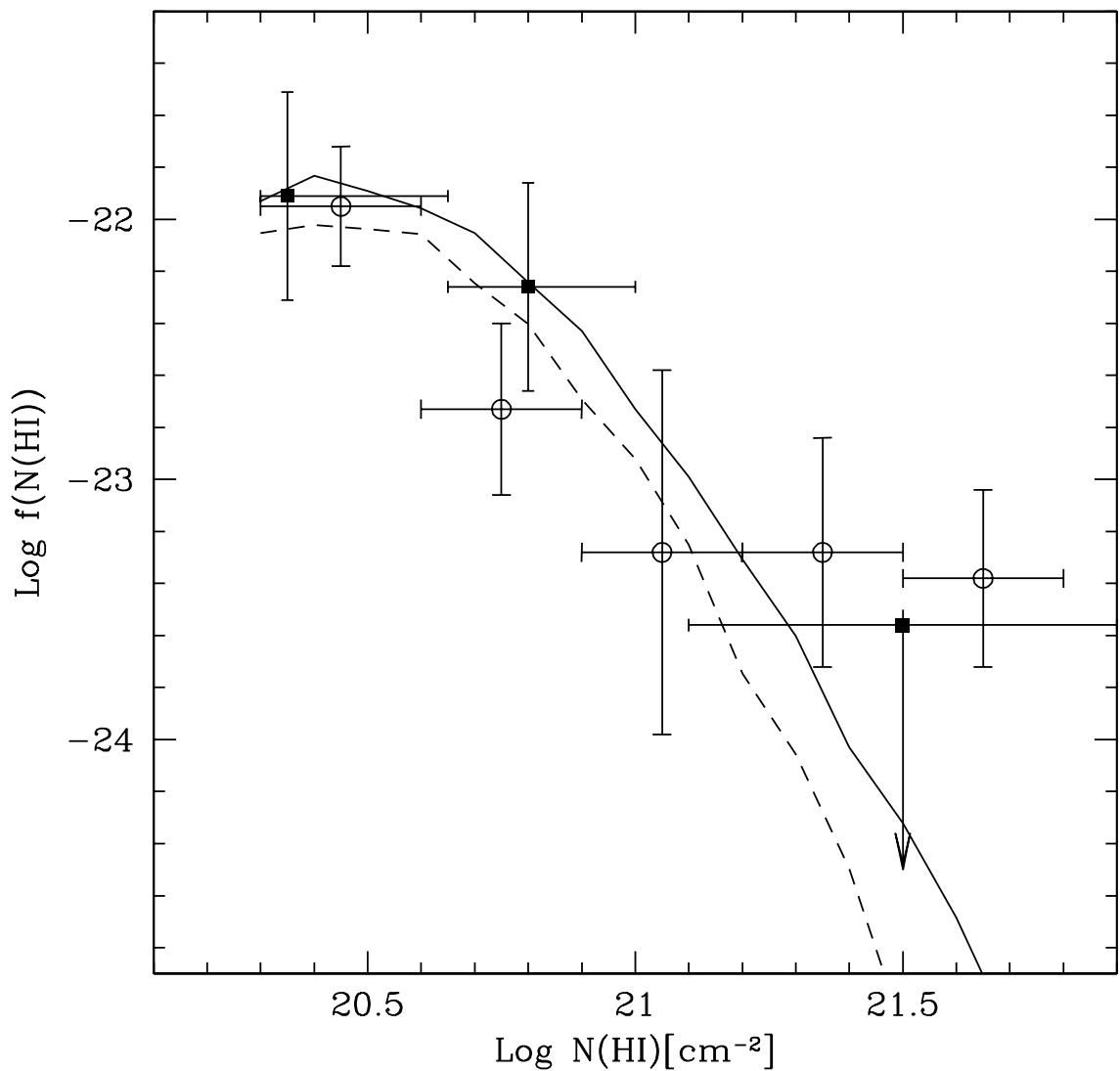}
\caption{
\ion{H}{1} column density distribution $f(N_{\rm HI})$ of DLA galaxies 
at $z=1$ without selection bias (solid line). 
We also show $f(N_{\rm HI})$ with the selection effect caused by the
 limitation of angular size $\theta_{\rm th} = 1$ [arcsec] (dashed line). 
The square symbols with crosses are the observed data shown  
\citep{SW00} (closed square).
Also shown  for another observation at 
$ \langle z \rangle = 0.78$ 
\citep{RT00} (open circle).      
 } \label{fig:corrfil}
\end{figure*}

\end{document}